\documentclass{article}

\def\isICASSPversion{} % a switch for the preamble
\def\isARXIVversion{} % turns on appendix and such
\ifdefined\isICASSPversion
\def\ARXIVurl{\url{https://arxiv.org/abs/XYZ}}
\usepackage{spconf}
\ninept
\else
\usepackage[
hyperref=true,
url=true,
style=numeric-comp,
maxbibnames=100,
backend=bibtex8
]{biblatex}
\fi

\ifdefined\isARXIVversion
\usepackage{appendix}
\fi

\usepackage[T1]{fontenc}
\usepackage{amsmath,amsfonts,amssymb,amsthm}
\usepackage{pifont}

\usepackage{booktabs}

\usepackage{cite}
\usepackage{doi}

\usepackage{hyperref}
\hypersetup{
  colorlinks = true,
  allcolors = blue,
}
\usepackage[capitalise,nameinlink]{cleveref}

\usepackage{graphicx}
\usepackage{tikz}
\usetikzlibrary{matrix,arrows,shapes,positioning,fit,calc,decorations.pathmorphing,intersections,shapes.misc}
\usepackage{arrayjobx}
\usepackage{xifthen}
\newcommand{\unionclip}[3]{% number of clippings, clipping commands (connect with &), draw commands
\newarray\temp%
\readarray{temp}{#2}%
\foreach \x in {1,...,#1}%
{ \begin{scope}%
\pgfmathtruncatemacro{\xt}{\x}%
\temp(\xt)%
#3%
\end{scope}%
}%
\delarray\temp%
}

\usepackage{xcolor}
\definecolor{myYellow}{rgb}{1.0,1.0,0.898}
\definecolor{myYellowGreen}{rgb}{0.471,0.776,0.475}
\definecolor{myGreen}{rgb}{0.0,0.271,0.161}
\pgfdeclareverticalshading{YlGn}{100bp}{color(0bp)=(myYellow); color(50bp)=(myYellowGreen); color(100bp)=(myGreen)}

\definecolor{myDarkBlue}{rgb}{0.0,0.0,0.3}
\definecolor{myBlue}{rgb}{0.0,0.0,1.0}
\definecolor{myWhite}{rgb}{1.0,1.0,1.0}
\definecolor{myRed}{rgb}{1.0,0.0,0.0}
\definecolor{myDarkRed}{rgb}{0.5,0.0,0.0}
\pgfdeclareverticalshading{seismic}{60bp}{color(0bp)=(myDarkBlue); color(15bp)=(myBlue); color(30bp)=(myWhite); color(45bp)=(myRed); color(60bp)=(myDarkRed)}

\usepackage[outline]{contour}
\contourlength{2.0pt}

\usepackage{pgfplots}
\pgfplotsset{compat=1.17}
\usepgfplotslibrary{groupplots}
\usepgfplotslibrary{fillbetween}

\pgfplotscreateplotcyclelist{mycyclelist}{%
  {red, mark=o, mark options={solid}, solid, thick},
  {blue, mark=x, mark options={solid}, dashed, thick},
  {violet, mark=*, mark options={solid}, loosely dotted, thick},
  {orange, mark=o, mark options={solid}, dashed, thick},
  {black, mark=x, mark options={solid}, solid, thick}
}
\pgfplotsset{cycle list name=mycyclelist}

\newtheorem{prop}{Proposition}
\newtheorem{theorem}{Theorem}

\newtheoremstyle{myremark}
{\topsep} % space above
{\topsep} % space below
{\normalfont} % body font
{} % indent amount
{\bfseries} % theorem head font
{.} % punctuation after theorem head
{5pt plus 1pt minus 1pt} % space after theorem head
{\thmname{#1}\thmnumber{ #2}\thmnote{ (#3)}} % theorem head spec

\theoremstyle{myremark}

\theoremstyle{definition}

\crefmultiformat{prop}{Propositions~#2#1#3}{ and~#2#1#3}{, #2#1#3}{ and~#2#1#3}
\crefmultiformat{theorem}{Theorems~#2#1#3}{ and~#2#1#3}{, #2#1#3}{ and~#2#1#3}
\crefmultiformat{lemma}{Lemmas~#2#1#3}{ and~#2#1#3}{, #2#1#3}{ and~#2#1#3}
\crefmultiformat{coro}{Corollaries~#2#1#3}{ and~#2#1#3}{, #2#1#3}{ and~#2#1#3}
\crefmultiformat{problem}{Problems~#2#1#3}{ and~#2#1#3}{, #2#1#3}{ and~#2#1#3}
\crefmultiformat{assump}{Assumptions~#2#1#3}{ and~#2#1#3}{, #2#1#3}{ and~#2#1#3}

\crefmultiformat{defn}{Definitions~#2#1#3}{ and~#2#1#3}{, #2#1#3}{ and~#2#1#3}
\crefrangeformat{defn}{Definitions.~#3#1#4,#5#2#6}

\crefmultiformat{property}{Properties~#2#1#3}{ and~#2#1#3}{, #2#1#3}{ and~#2#1#3}
\crefrangeformat{property}{Properties~#3#1#4,#5#2#6}

\newcommand{\eg}{\textit{e.g.}}

\newcommand{\bc}{\ensuremath{\mathbf{c}}}

\newcommand{\be}{\ensuremath{\mathbf{e}}}
\newcommand{\boldf}{\ensuremath{\mathbf{f}}}
\newcommand{\bu}{\ensuremath{\mathbf{u}}}
\newcommand{\bv}{\ensuremath{\mathbf{v}}}

\newcommand{\bx}{\ensuremath{\mathbf{x}}}

\newcommand{\bA}{\ensuremath{\mathbf{A}}}

\newcommand{\bbR}{\ensuremath{\mathbb{R}}}

\newcommand{\cC}{\ensuremath{\mathcal{C}}}
\newcommand{\cD}{\ensuremath{\mathcal{D}}}

\newcommand{\del}{\ensuremath{\partial}}

\DeclareMathOperator{\Ima}{Im}
\DeclareMathOperator{\rank}{rank}

\title{Hodgelets: Localized Spectral Representations \\of Flows on Simplicial Complexes}
\name{%
T.~Mitchell~Roddenberry$^*$, Florian~Frantzen$^\dagger$, Michael~T.~Schaub$^\dagger$, Santiago~Segarra$^*$%
\thanks{This work was partially supported by USA NSF under award CCF-2008555. FF and MTS acknowledge partial support from the Excellence Strategy of the Federal Government and the Länder in Germany, and the NRW Rückkehrprogramm.}
}

\address{$^*$Rice University, Dept. of Electrical and Computer Engineering, Houston, TX, USA \\
$^\dagger$RWTH Aachen University, Dept. of Computer Science, Aachen, Germany}

\begin{document}
\maketitle
\begin{abstract}
We develop wavelet representations for edge-flows on simplicial complexes, using ideas rooted in combinatorial Hodge theory and spectral graph wavelets.
We first show that the Hodge Laplacian can be used in lieu of the graph Laplacian to construct a family of wavelets for higher-order signals on simplicial complexes.
Then, we refine this idea to construct wavelets that respect the Hodge-Helmholtz decomposition.
For these \emph{Hodgelets}, familiar notions of curl-free and divergence-free flows from vector calculus are preserved.
We characterize the representational quality of our Hodgelets for edge flows in terms of frame bounds and demonstrate the use of these spectral wavelets for sparse representation of edge flows on real and synthetic data.
\end{abstract}
\section{Introduction}

There has been substantial interest in graph-based techniques to understand data with a complex relational structure~\cite{Strogatz:2001,Newman:2010,Jackson:2010}, with applications ranging from biology~\cite{Garroway:2008} to system robustness~\cite{Holme:2002}.
In this context, graph signal processing (GSP) has proven to be a useful way to understand the processing of signals defined on graphs, leveraging ideas from both signal processing and graph theory~\cite{Shuman:2013}.
The primary focus of GSP has been on signals supported on the \emph{nodes} of a graph.
For such signals, the graph Laplacian and adjacency matrix are natural shift operators, from which we can define notions of filtering and Fourier transformations~\cite{Shuman:2013}.

However, there has been a recent flurry of interest in studying \emph{flows} on the edges of graphs and simplicial complexes~\cite{Schaub:2018,Jia:2019,Barbarossa:2020a,Barbarossa:2020b,Yang:2021,Schaub:2021}, which can be used to model the motion of mass, energy, or information.
Since flows carry a natural orientation that does not arise when studying signals on the nodes of a graph, recent works have leveraged tools from algebraic topology~\cite{Hatcher:2005} and discrete exterior calculus~\cite{Hirani:2003} to form appropriate Laplace operators that respect the orientation of edge flows.
This approach has allowed for the study of edge flows through the lens of the celebrated Hodge-Helmholtz decomposition~\cite{Schaub:2018,Jia:2019,Barbarossa:2020a,Barbarossa:2020b}.
This viewpoint has even been leveraged to define neural network architectures for edge flows~\cite{Roddenberry:2019,Ebli:2020,Bunch:2020,Roddenberry:2021,Bodnar:2021}.

In the literature thus far, the primary focus has been on understanding filtering and signal representation in the spatial and Fourier domains, where we take the Fourier modes to be the eigenvectors of a suitably defined Laplacian.
However, just as in classical signal processing, the Fourier modes are highly \emph{delocalized}.
That is to say, the support of a Fourier mode is typically not restricted to one small region of the graph.
In GSP, this has motivated the development of \emph{spectral graph wavelets}~\cite{Hammond:2011,Shuman:2015}, which proposes to use a dictionary of atoms for signal representation that is localized in both the spatial and frequency domains.
Here, we introduce a family of \emph{wavelets for edge flows}, seeking to balance localization in the spatial and frequency domains, while also respecting the Hodge decomposition.

\noindent\textbf{Contributions and outline.}
We consider the design of \emph{spectral wavelets} for edge flows on simplicial complexes.
In particular, we discuss how the orthogonal decomposition of the space of edge flows in terms of the Hodge Laplacian can be leveraged to design interpretable wavelets that yield high-quality sparse and localized representations of edge flows.

We begin by discussing preliminaries in graph signal processing for edge flows in \cref{sec:background}.
Then, we propose a simple construction for spectral graph wavelets based on previous literature in \cref{sec:model}, as well as a modification that respects the Hodge decomposition.
Theoretical properties of both models are considered in \cref{sec:theory}. 
In particular, we state frame bounds for both models, in terms of the family of spectral kernels used in their definition.
Finally, we demonstrate the utility of our constructions for sparse representation and flow clustering on real and synthetic data in \cref{sec:experiments}.

\section{Notation and Background}\label{sec:background}

For a positive integer $N$, we denote the set of integers $\{1,2,\ldots,N\}$ by $[N]$.
We use $\cong$ to denote isomorphism between vector spaces, and $\oplus$ to denote the orthogonal direct sum of vector spaces.
For a linear operator $\bA$ between two vector spaces, we denote the set of eigenvalues of $\bA$ by $s(\bA)$.

\noindent\textbf{Simplicial complexes and the Hodge Laplacian.}
We consider data supported on \emph{(abstract) simplicial complexes}, which generalize graphs to allow for higher-order connectivity.
An {(abstract) simplicial complex} $X$ is a finite collection of finite sets that is closed under restriction: that is to say, for any $\sigma$ in $X$, all nonempty subsets of $\sigma$ are also contained in $X$.
We call the elements of $X$ \emph{simplices} and denote by $X_k$ the set of all simplices in $X$ with cardinality $k+1$, also referred to as \emph{$k$-simplices}.
In particular, $X_0$ is the set of all singleton sets in $X$, $X_1$ is the set of all simplices with cardinality $2$, and so on.
Grounded in our intuition for graphs, we call $X_0$ the set of \emph{nodes} in $X$, $X_1$ the set of \emph{edges} in $X$, and $X_2$ the set of \emph{triangles}.

We identify the set $X_k$ with the integers $[N_k]$, and denote the cardinality of $X_k$ by $N_k$.
By convention, we label the nodes with $1, 2, \ldots, N_0$.
Further, we assign to each $k$-simplex an \emph{orientation}\footnote{The choice of orientation is arbitrary and distinct from the notion of direction, e.g., in a directed graph. See \cite{Schaub:2020,Lim:2020} for details.}, or a canonical ordering, following the ordering induced by the node labeling, \eg, a triangle $\{i,j,k\}$ is given the orientation $[i,j,k]$, where $i<j<k$.
Given this reference orientation, we define the space of \emph{$k$-cochains}, denoted by $\cC^k(X)$, as the vector space of functions from the oriented simplices in $X_k$ to $\bbR$.
One can check that for each $k$, $\cC^k(X)$ is naturally isomorphic to $\bbR^{N_k}$.
For a given $\cC^k(X)$, we take as an orthonormal basis for that space the set of functions $\{\be_\sigma\}_{\sigma\in X_k}$ taking unit value on each oriented $k$-simplex $\sigma$, and zero elsewhere.
Of particular interest is the space $\cC^1(X)$, which models flows on the edges of a graph or simplicial complex~\cite{Schaub:2018,Jia:2019,Barbarossa:2020a,Barbarossa:2020b}.

The spaces $\cC^k(X)$ are related by the set of \emph{incidence matrices}.
The 2nd incidence matrix $\del_2:\cC^2(X)\to\cC^1(X)$ is a linear map defined over the standard orthonormal basis for $\cC^2(X)$ as follows: $\del_2\be_{[i,j,k]}=\be_{[i,j]}+\be_{[j,k]}-\be_{[i,k]}$.
Similarly, the 1st incidence matrix $\del_1:\cC^1(X)\to\cC^0(X)$ is the familiar node-edge incidence matrix, defined according to $\del_1\be_{[i,j]}=\be_{[j]}-\be_{[i]}$.
We also call the matrices $\del_2,\del_1$ the \emph{boundary maps}, since they map each $k$-simplex to the $k-1$-simplices on its boundary, as well as obeying the important property $\del_1\del_2=0$.
The following well-known result characterizes the space $\cC^1(X)$ in terms of these maps.

\begin{theorem}[Hodge Decomposition]\label{thm:hodge}
  Let $X$ be a finite simplicial complex.
  Define the first Hodge Laplacian as $\Delta_1=\del_{2}\del_{2}^\top+\del_1^\top\del_1$.
  The vector space $\cC^1(X)$ can be written as the direct sum of orthogonal subspaces:
  \begin{equation}
    \cC^1(X) \cong \Ima(\del_2)
    \oplus\Ima(\del_1^\top)
    \oplus\ker(\Delta_1).
  \end{equation}
\end{theorem}
See \cite{Lim:2020} for the proof.
Viewing the standard graph Laplacian $\Delta_0=\del_1\del_1^\top$ as an operator that measures smoothness for node signals based on their edgewise incidence, the Hodge Laplacian $\Delta_1$ similarly measures ``smoothness'' for $1$-cochains based on nodewise and trianglewise incidence.
For convenience, we define the \emph{upper Hodge Laplacian} $\Delta_1^U=\del_2\del_2^\top$ and the \emph{lower Hodge Laplacian} $\Delta_1^L=\del_1^\top\del_1$, so that $\Delta_1=\Delta_1^U+\Delta_1^L$.

\noindent\textbf{Spectral graph wavelets.}
A key component of many modern signal processing and machine learning tasks is the choice of a proper representation for the data.
In graph signal processing, this often amounts to using a spatial representation or a frequency domain representation.
In time-domain signal processing, one can construct a dictionary of \emph{wavelets} that interpolates between these two extremes~\cite{Mallat:1999}.
Similarly, in \cite{Hammond:2011,Shuman:2015}, methods to construct \emph{spectral graph wavelets} are proposed, in which a family of kernel functions $\{g_m:\bbR\to\bbR\}_{m\in\Gamma}$ for some index set $\Gamma$ is applied to the standard basis for $\cC^0(X)$ on a graph $X$ to yield a dictionary for localized, bandlimited representations of graph signals as follows:
\begin{equation}
    \cD = \left\{\psi_{j,m}=g_m(\Delta_0)\be_j:m\in\Gamma,j\in [N_0]\right\},
\end{equation}
where $\Delta_0$ indicates the graph Laplacian.

\section{An Interpretable Spectral Wavelet Model}\label{sec:model}

\begin{figure}
    \centering
    \resizebox{\linewidth}{!}{\begin{tikzpicture}[shading=seismic]
  \tikzstyle{every node}=[font=\Large]
  \begin{groupplot}[
    group style={
      group size=2 by 2,
      group name=myplots,
      horizontal sep=2em,
      vertical sep=1em,
    },
    hide axis,
    enlargelimits=false,
    ]
    
    \nextgroupplot[title={}]
    \addplot graphics[xmin=0,xmax=1,ymin=0,ymax=1] {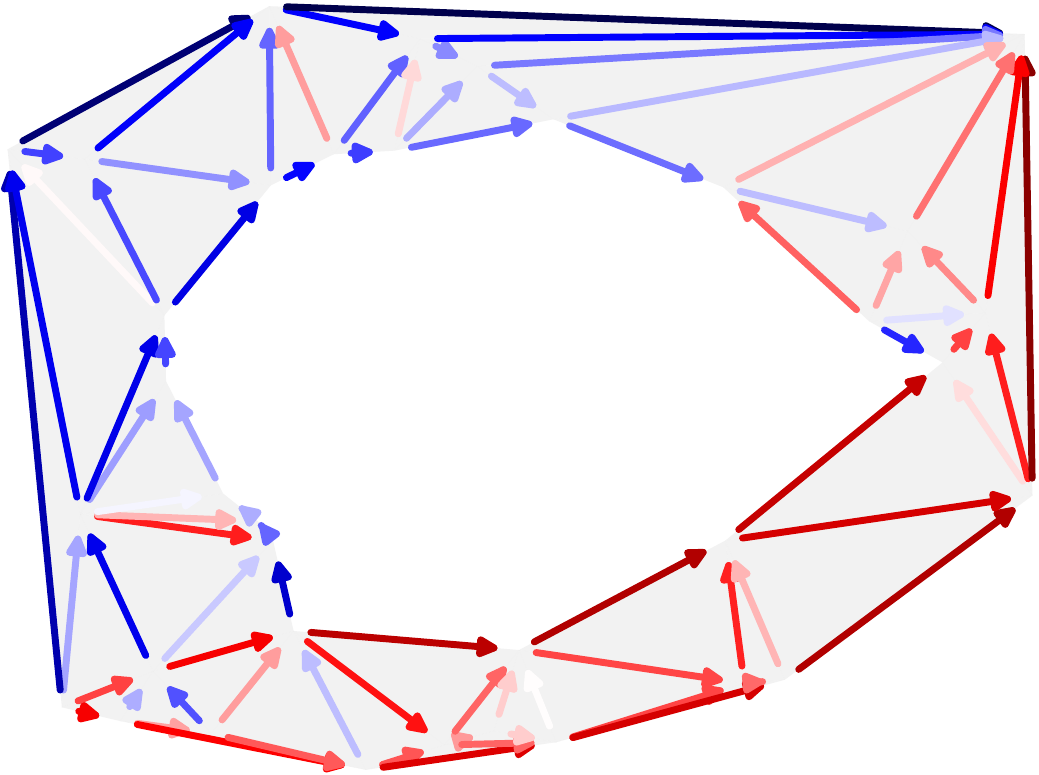};
    \nextgroupplot[title={}]
    \addplot graphics[xmin=0,xmax=1,ymin=0,ymax=1] {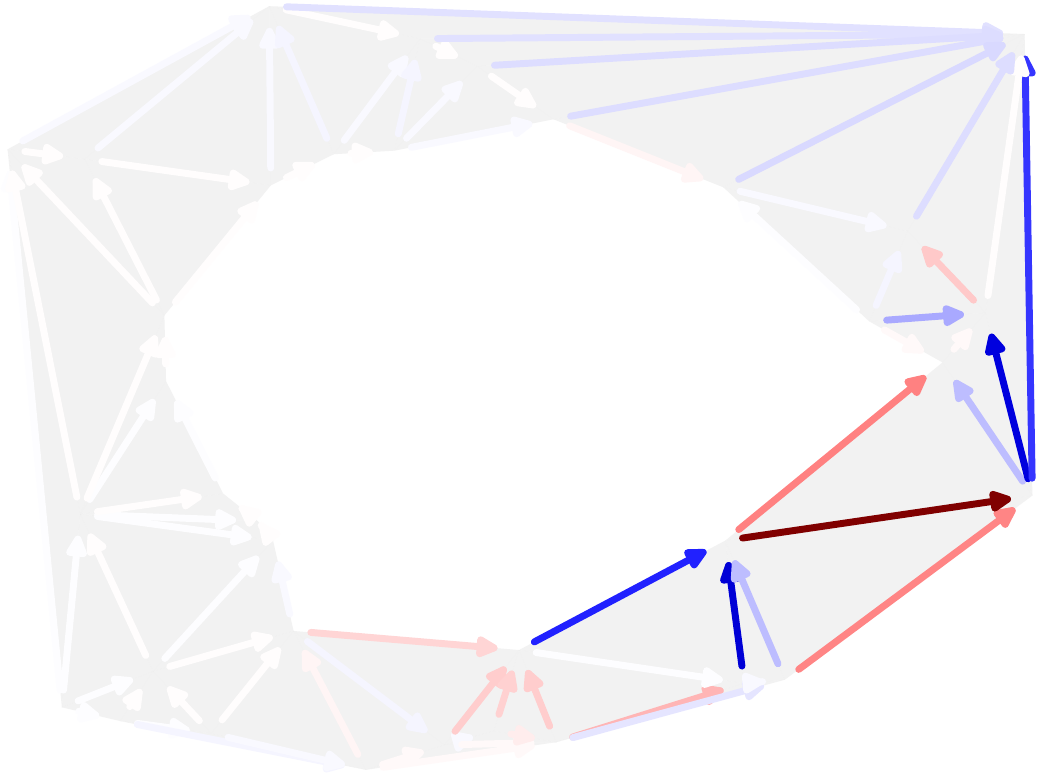};
    \nextgroupplot[title={}]
    \addplot graphics[xmin=0,xmax=1,ymin=0,ymax=1] {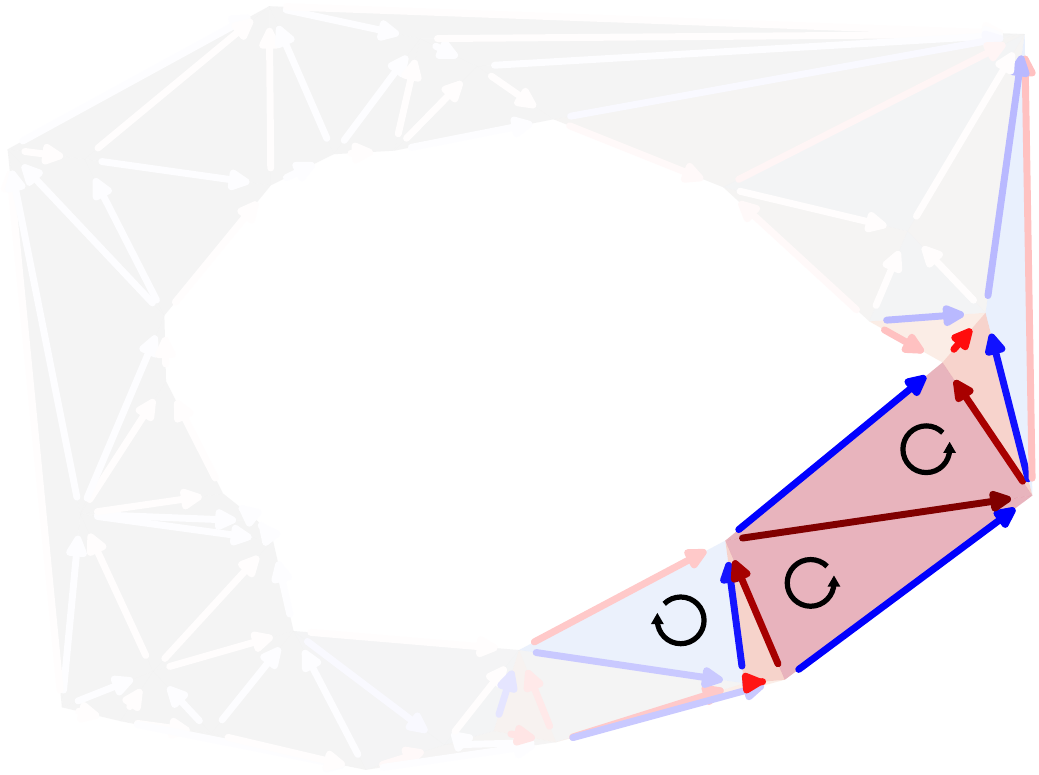};
    \nextgroupplot[title={}]
    \addplot graphics[xmin=0,xmax=1,ymin=0,ymax=1] {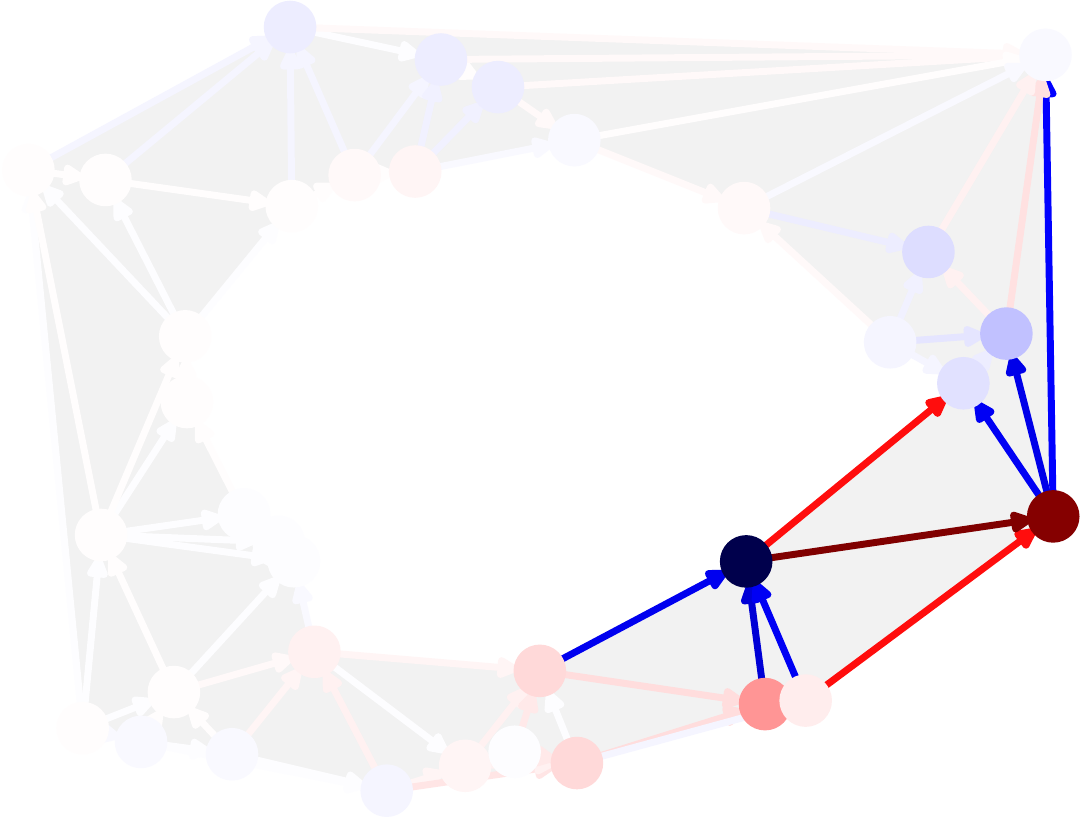};
    
  \end{groupplot}
  
  % axis
  \foreach \plt/\lab in {c1r1/a,c2r1/b,
  c1r2/c1,c2r2/c2} {
    \node[anchor=north west] at (myplots \plt.north west) {(\lab)};
  }

  % colorbar
  \shadedraw[color=black, very thin, shading angle=0] ($(myplots c2r1.north east)+(0.5,-2.0)$) rectangle ($(myplots c2r2.south east)+(1.0,2.0)$);
  \node[anchor=north west] at ($(myplots c2r1.north east)+(1.0,-2.0)$) {$+1.0$};
  \node[anchor=west] at ($(myplots c2r1.north east)!0.5!(myplots c2r2.south east)+(1.0,0)$) {$\phantom{+}0.0$};
  \node[anchor=south west] at ($(myplots c2r2.south east)+(1.0,2.0)$) {$-1.0$};
  
\end{tikzpicture}}
    \caption{
    Dictionary atoms for representing flows on simplicial complexes.
    All signals are scaled to have unit $\ell_\infty$-norm.
    Edge orientations are indicated by the direction of the arrowheads.
    (a) An eigenvector in the nullspace of the Hodge Laplacian $\Delta_1$.
    (b) Jointly designed wavelet based on $\Delta_1$.
    (c1) Upper wavelet based on $\Delta_1^U$. Triangles are colored according to the $2$-cochain $\bc$ such that the wavelet atom is equal to $\del_2\bc$, with orientation of the flow induced by each triangle indicated by arrows (space permitting).
    (c2) Lower wavelet based on $\Delta_1^L$. Nodes are colored according to the $0$-cochain $\bx$ such that the wavelet atom is equal to $\del_1^\top\bx$.
    }
    \label{fig:localization}
\end{figure}
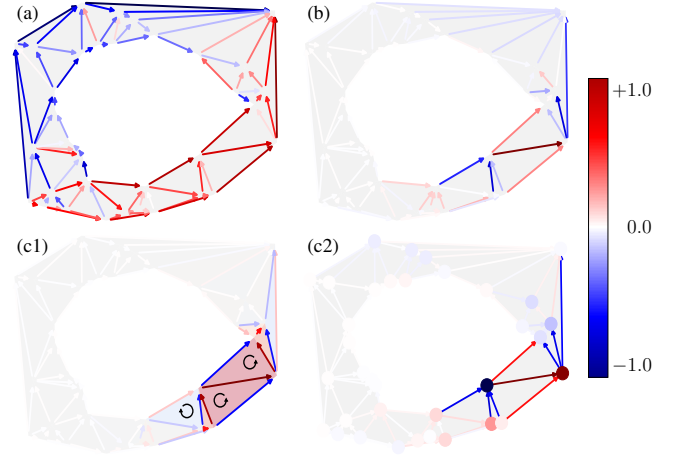

Much like in graph signal processing, we seek useful representations of edge flows that balance spatial localization and bandlimitedness in the frequency domain.
As done by \cite{Yang:2021}, we can define a \emph{simplicial Fourier transform} by projecting an edge flow onto each eigenvector of $\Delta_1$.
To illustrate this, we construct a simplicial complex by picking $40$ points randomly distributed in the unit square, then taking their Delaunay triangulation~\cite{Delaunay:1934}.
We then create one hole in this simplicial complex, and construct the Hodge Laplacian $\Delta_1$ for the remaining structure.
We illustrate this complex as well as one of the eigenvectors in its nullspace in \cref{fig:localization}~(a). 
We see that the flow ``wraps around'' the hole of the complex, an inherently global phenomenon, but does not have \emph{localized} support on the complex.
In the ensuing discussion, we aim to construct dictionaries for edge flows that find a balance in this local-global tradeoff.

\noindent\textbf{Jointly designed wavelets.}
As the Hodge Laplacian generalizes the graph Laplacian for higher-order signals, it is natural to apply the methods of \cite{Hammond:2011,Shuman:2015} for constructing graph wavelets by substituting the Hodge Laplacian for the graph Laplacian.
In that direction, let $\{\be_j\}_{j=1}^{N_1}$ be the standard orthonormal basis for $\cC^1(X)$, and let $\{g_m\}_{m=1}^M$ be a set of continuous, non-negative functions on the real line.
For each $j\in[N_1], m\in[M]$, we define the atom $\psi_{j,m}$ by applying the polynomial $g_m(\Delta_1)$ to $\be_j$, where we have implicitly labeled the oriented $1$-simplices with the integers $[N_1]$.
That is, if the Hodge Laplacian $\Delta_1$ admits an eigendecomposition $\{(\lambda_i,\bv_i)\}_{i=1}^{N_1}$,
\begin{equation}\label{eq:atoms-joint}
  \psi_{j,m} =
  g_m(\Delta_1)\be_j =
  \left(\sum_{i=1}^{N_k}g_m(\lambda_i)\bv_i\bv_i^\top\right)\be_j.
\end{equation}
Since this construction uses the sum of the upper and lower Hodge Laplacians and thus both components of the Laplacian, we refer to such atoms as \emph{joint Hodgelets}.

We illustrate this approach in \cref{fig:localization}~(b), using the same simplicial complex as before.
We construct wavelets using the log-scaled Hann kernels of \cite{Shuman:2015}, and illustrate a single wavelet atom.
Note that this atom is spatially localized, due to the spectral kernel being well-approximated by a low-order polynomial of the Hodge Laplacian.

\noindent\textbf{Separately designed wavelets.}
The approach of directly applying the wavelet construction of~\cite{Hammond:2011,Shuman:2015} using the Hodge Laplacian presents some shortcomings.
Importantly, it fails to differentiate between the spectral features of the Hodge Laplacian due to the upper and lower components being present in each atom.
To address this issue, we treat each subspace of the Hodge decomposition~[cf. \cref{thm:hodge}] separately, by defining \emph{upper} and \emph{lower} wavelets for $\cC^1(X)$.
A similar approach was taken in the design of filters for signals on simplicial complexes by \cite{Yang:2021}.

As before, let $\{\be_j\}_{j=1}^{N_1}$ be the standard orthonormal basis for $\cC^1(X)$, and let $\{g_m^U\}_{m=1}^{M_U}, \{g_m^L\}_{m=1}^{M_L}$ be sequences of continuous functions on the real line.
For each $j\in[N_k], m\in[M_U], m'\in[M_L]$, define
\begin{equation}\label{eq:atoms-upperlower}
  \begin{aligned}
    \psi_{j,m}^U &= g_m^U(\Delta_1^U)\be_j \\
    \psi_{j,m'}^L &= g_{m'}^L(\Delta_1^L)\be_j,
  \end{aligned}
\end{equation}
where $g_m^U(\Delta_1^U), g_m^L(\Delta_1^L)$ are defined via the functional calculus as before.
The set $\{\psi_{j,m}^U\}$ forms what we call the \emph{upper atoms}, and similarly $\{\psi_{j,m}^L\}$ forms the set of \emph{lower atoms} for $\cC^1(X)$.
Since this construction separates the upper and lower components of the Hodge Laplacian, we refer to such atoms as \emph{separate Hodgelets}.

By separately treating the upper and lower Hodge Laplacian, we can construct atoms with greater interpretability than those designed jointly.
In particular, we can show the following result:
\begin{prop}\label{prop:div-curl-wavelets}
  Suppose $g_m^U$ and $g_{m'}^L$ are kernels that take value $0$ at $0$.
  Then, for all $j,j'\in [N_1]$,
  \begin{equation}
      \psi_{j,m}^U\in\Ima(\del_2) \qquad \text{and} \qquad \psi_{j',m'}^L\in\Ima(\del_1^\top).
  \end{equation}
\end{prop}
Thus, the upper wavelets are dictated by the boundaries of $2$-simplices (triangles), and the lower wavelets are dictated by the coboundaries of $0$-simplices (nodes).
\ifdefined\isARXIVversion
%% you are in the ArXiV version: the proof is in the appendix
We leave the proof to \cref{app:div-curl}.
\else
%% you are not in the ArXiV version: omit the proof and reference the ArXiV version
We omit the proof for space reasons.%
\footnote{Proofs can be found at \ARXIVurl.}
\fi
Indeed, \cref{prop:div-curl-wavelets} reflects the properties of \cite[Theorem~3]{Lessig:2021}, in which curl and divergence wavelets are constructed for differential forms in Euclidean space.

We illustrate this in \cref{fig:localization}~(c1,c2), by plotting wavelet atoms with the same kernels as the joint wavelet in \cref{fig:localization}~(b), except with a separated construction.
One can see that there is a clear distinction between the upper wavelet (c1) which corresponds to a curl around triangles, and the lower wavelet (c2) which corresponds to the gradient of a node signal.

\section{Frame Bounds on Dictionaries}\label{sec:theory}

In signal processing on graphs, the graph Fourier transform has the appealing property of being an orthogonal transform, thus preserving the norm of the signal it acts upon.
Since wavelet dictionaries are typically overcomplete in their construction, we do not have orthogonality, but rather have \emph{frame bounds} for the dictionary.
For a Hilbert space $V$, a dictionary of vectors $\cD$ with at most countably many elements is said to be an $(A,B)$-frame with $0 \leq A \leq B < \infty$ if for all $v\in V$, we have
\begin{equation}
    A\|\bv\|^2 \leq \sum_{\psi\in\cD}|\langle\psi,\bv\rangle|^2 \leq B\|\bv\|^2.
\end{equation}
If $A=B$, we say that $\cD$ forms a \emph{tight frame}.
We allow for the case where $A=0$, in which case $\cD$ is a degenerate frame.
The frame bounds of a dictionary dictate its representational quality, as well as the performance of reconstruction algorithms~\cite{Mallat:1999}.
Moreover, if $A=B=1$, the coefficients $|\langle\psi,v\rangle|^2$ are analogous to the spectrogram representation of a signal~\cite{Shuman:2015}.
Here, in the same vein as \cite{Hammond:2011,Shuman:2015}, we characterize the frame bounds for both Hodgelet constructions in terms of the spectral properties of the kernels $g_m$.

\noindent\textbf{Joint wavelets.}
Given that the jointly designed wavelets are a direct adaptation of those proposed in \cite{Hammond:2011,Shuman:2015}, we can show a similar result in our context:
\begin{theorem}\label{thm:tight-frame-sufficient}{{\normalfont (based on \cite{Hammond:2011,Shuman:2015})}}
  Let $\{\be_j\}_{j=1}^{N_1}$ be the standard orthonormal basis of $\cC^1(X)$,
  and let $\{g_m\}_{m=1}^{M}$ be continuous non-negative functions on the real line.
  Let $\cD$ be the dictionary of atoms defined by \eqref{eq:atoms-joint}, and define
  \begin{equation}
    G(\lambda) = \sum_{m=1}^M|g_m(\lambda)|^2.
  \end{equation}
  Then, $\cD$ forms an $(A,B)$-frame for $\cC^1(X)$, where
  \begin{equation}
    A = \min_{\lambda\in s(\Delta_1)}G(\lambda) \quad B = \max_{\lambda\in s(\Delta_1)}G(\lambda).
  \end{equation}
\end{theorem}
We omit the proof, as it directly mirrors that of \cite{Shuman:2015}.
In particular, if $G$ is constant on $s(\Delta_0)$, then $\cD$ is a tight frame.

\noindent\textbf{Separate wavelets.}
We now state frame bounds for the separate Hodgelet construction, keeping the mutual orthogonality of the upper and lower Hodge Laplacians in mind.
\begin{theorem}\label{thm:tight-frame-hodge}
  Let $\{\be_j\}_{j=1}^{N_1}$ be the standard orthonormal basis for $\cC^1(X)$,
  and let $\{g_m^U\}_{m=1}^{M_U}, \{g_m^L\}_{m=1}^{M_L}$ be collections of continuous non-negative functions on the real line.
  Let $\cD$ be the dictionary of separate Hodgelets defined by \eqref{eq:atoms-upperlower},
  and define
  \begin{equation}\label{eq:partition-of-unity}
    G(\mu,\nu) = \sum_{m=1}^{M_U}|g_m^U(\mu)|^2 + \sum_{m=1}^{M_L}|g_m^L(\nu)|^2
  \end{equation}
  Then, $\cD$ forms an $(A,B)$-frame for $\cC^1(X)$, where
  \begin{equation}\label{eq:frame-bound-defn}
    \begin{aligned}
      A &= \min\left\{\min_{\mu\in s(\Delta_1^U)}G(\mu,0),
      \min_{\nu\in s(\Delta_1^L)}G(0,\nu)\right\} \\
      B &= \max\left\{\max_{\mu\in s(\Delta_1^U)}G(\mu,0),
      \max_{\nu\in s(\Delta_1^L)}G(0,\nu)\right\}.
    \end{aligned}
  \end{equation}
\end{theorem}
\ifdefined\isARXIVversion
%% you are in the ArXiV version: the proof is in the appendix
We leave the proof to \cref{app:tight-frame-sep}.
\fi
By \cref{thm:tight-frame-hodge}, we see that the frame bounds are determined by the quality of the kernels for the upper and lower parts of the spectrum independently.
In particular, if $G(\cdot,0)$ and $G(0,\cdot)$ are constant on $s(\Delta_1^U)$ and $s(\Delta_1^L)$, respectively, then the dictionary forms a tight frame.

In the context of \cref{prop:div-curl-wavelets}, \cref{thm:tight-frame-hodge} indicates that we can construct norm-preserving representations of edge flows that are also interpretable in terms of harmonic flows in $\ker(\Delta_1)$, curl flows in $\Ima(\del_2)$, and divergence flows in $\Ima(\del_1^\top)$.
This is aligned with the development in~\cite{Yang:2021}, where edge flow filters were designed using $\Delta_1^L$ and $\Delta_1^U$ separately, rather than the total Hodge Laplacian $\Delta_1$.

\section{Experiments}\label{sec:experiments}

\begin{figure}
  \centering
  \resizebox{\linewidth}{!}{\begin{tikzpicture}[shading=seismic]
  \tikzstyle{every node}=[font=\Large]
  \pgfplotstableset{col sep=comma}
  \begin{groupplot}[
    group style={
      group size=2 by 2,
      group name=myplots,
      horizontal sep=6em,
    },
    ylabel near ticks,
    ]

    \nextgroupplot[
    domain=-2:2,
    xmin=-2, xmax=2,
    ymin=-2, ymax=2,
    view={0}{90},
    axis background/.style={fill=white}
    ]
    % \clip (0,0,0) -| (2,2,0) -| (0,0,0) -| (-2,-2,0) -| cycle;
    \unionclip{2}
    {\clip (pi/4,pi/4) circle (0.7);&\clip (-pi/4,-pi/4) circle (0.7);}
    {
    \addplot3[domain=-2:2, y domain=-2:2,
    blue,
    quiver={u={cos(deg(x+y))},v={sin(deg(x-y))},
      scale arrows=0.3},-stealth,samples=20]
    {0};
    }

    \nextgroupplot[
    xlabel={Error tol.},
    ylabel={Sparsity},
    xmode=log,
    legend pos=north east,
    ]
    
    \addplot+ table[x=err, y=fourier] {figs/hexgrid/sparsity.csv};
    \addlegendentry{Fourier}
    \addplot+ table[x=err, y=fourier_line] {figs/hexgrid/sparsity.csv};
    \addlegendentry{Fourier (LG)}
    \addplot+ table[x=err, y=joint] {figs/hexgrid/sparsity.csv};
    \addlegendentry{Joint}
    \addplot+ table[x=err, y=line] {figs/hexgrid/sparsity.csv};
    \addlegendentry{Joint (LG)}
    \addplot+ table[x=err, y=sep] {figs/hexgrid/sparsity.csv};
    \addlegendentry{Separated}

    \nextgroupplot[hide axis, enlargelimits=false,]
    \addplot graphics[xmin=0,xmax=1,ymin=0,ymax=1] {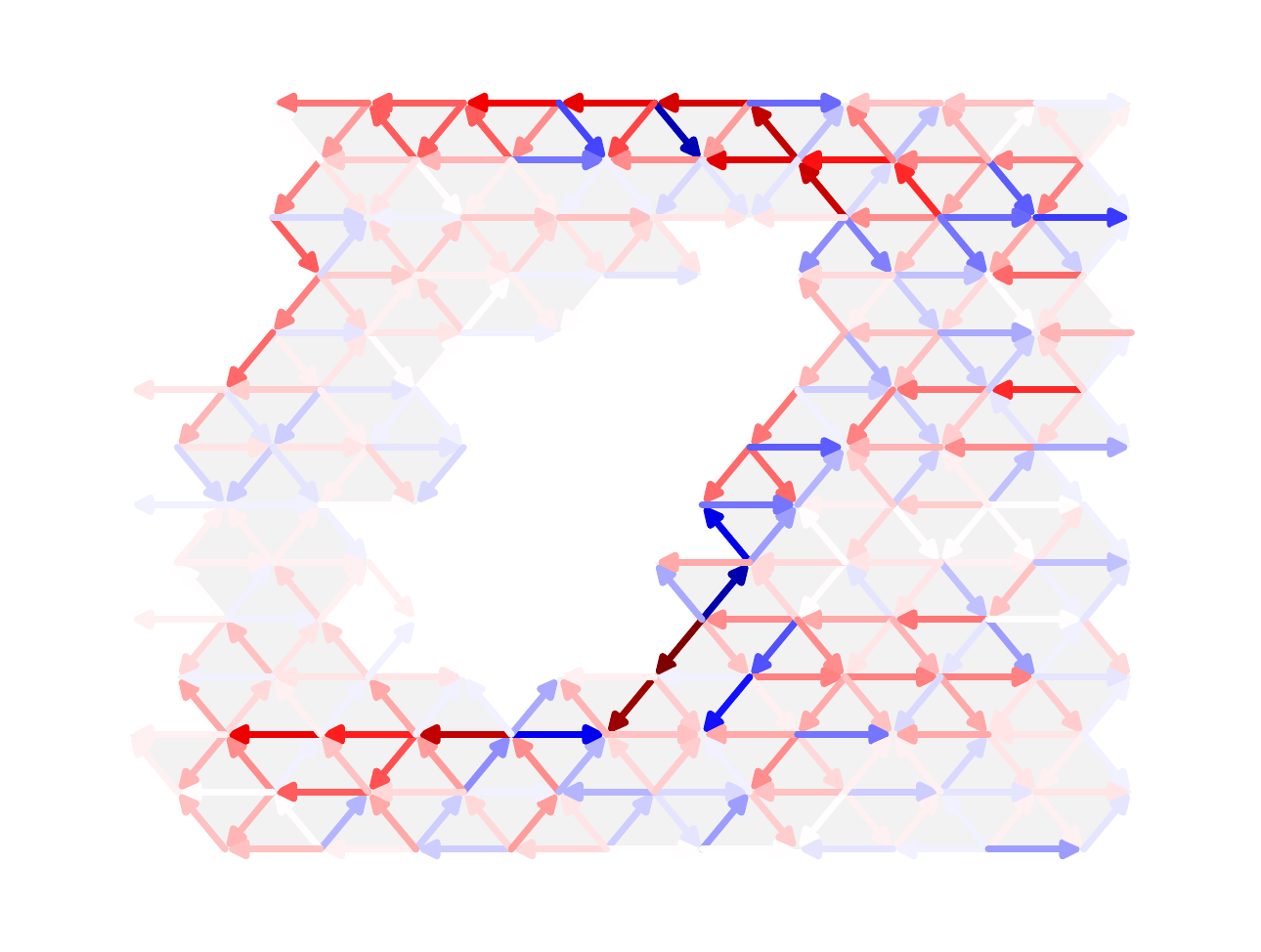};

    \nextgroupplot[hide axis, enlargelimits=false]
    
  \end{groupplot}
  
  % data table
  \node at (myplots c2r2) {%
    \LARGE
    \begin{tabular}{cc}
        \toprule
        Method & $L$ \\
        \midrule
        Standard & 0.076 \\
        Fourier & 0.204 \\
        Joint & 0.187 \\
        Separate & \textbf{0.495} \\
        \bottomrule
    \end{tabular}
  };
  
  % axis
  \foreach \plt/\lab in {c1r1/a,c2r1/b,c1r2/c,c2r2/d} {
    \node[anchor=north west,fill=white] at (myplots \plt.north west) {(\lab)};
  }
  
  % colorbar
  \shadedraw[color=black, very thin, shading angle=-90] ($(myplots c1r2.south west)+(0.5,0.25)$) rectangle ($(myplots c1r2.south east)+(-0.5,-0.25)$);
  \node[anchor=north west] at ($(myplots c1r2.south west)+(0.5,-0.25)$) {$-1.0$};
  \node[anchor=north] at ($(myplots c1r2.south west)!0.5!(myplots c1r2.south east)+(0,-0.25)$) {$0.0$};
  \node[anchor=north east] at ($(myplots c1r2.south east)+(-0.5,-0.25)$) {$+1.0$};
\end{tikzpicture}}
  \caption{% 
    Representation of flows on simplicial complexes with spectral wavelets.
    (a) Vector field $F(x,y)$.
    (b) Sparsity of representing a discretization of the vector field using different dictionaries as error tolerance increases. `LG' indicates the use of the linegraph Laplacian.
    (c) Sum of all buoy trajectories in the Global Drifter dataset around Madagascar. Signal is scaled to have unit $\ell_\infty$-norm.
    (d) Average maximum inner product $L$ between test set and centroids obtained via sparse $k$-means clustering using standard basis, Fourier basis, joint Hodgelets, and separate Hodgelets.
  }
  \label{fig:results}
\end{figure}
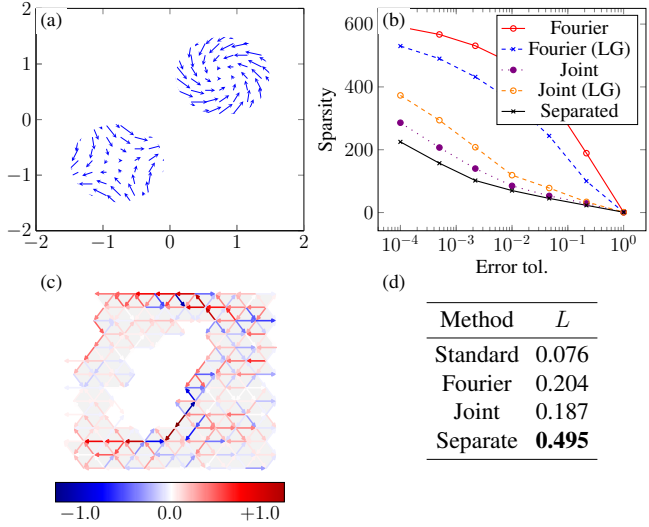

We demonstrate the utility of applying spectral wavelets based on the Hodge Laplacian for sparse, localized representations of flow data.
For all experiments,\footnote{Code is available at \url{https://www.git.roddenberry.xyz/hodgelets/}}
we take the spectral kernels $\{g_m\}_{m=1}^M$ to be the log-scaled Hann kernels proposed by \cite{Shuman:2015} with $R=3$, where $R$ dictates the degree of overlap between each kernel, and $M$ is chosen based on the particular task.
% \michael{maybe we can write out the formula for the Hann kernels somewhere (space permitting)? What R is would then also be more clear, supposedly} \mitch{The particular equation isn't too enlightening (see \cite{Shuman:2015} eq. 7-11). It really only helps to plot the kernel, but I'm not sure we have the space for that... I would say the space is better used by the figures we currently have.}\michael{Ok}

\noindent\textbf{Sparse representations.}
To illustrate the advantage of using wavelets based on the upper and lower Hodge Laplacians separately, we consider a vector field on $[-2,2]^2$, given by
\begin{equation}
  F(x,y) =
  \begin{cases}
    [\cos(x+y), \sin(x-y)] & (x,y)\in B_1\cup B_2, \\
    [0,0] & \text{otherwise},
  \end{cases}
\end{equation}
where we define $B_1, B_2$ to be closed balls of radius $0.7$ centered at $(\pm\pi/4,\pm\pi/4)$, respectively, as illustrated in \cref{fig:results}~(a).
We discretize $[-2,2]^2$ with a hexagonal grid, then construct a simplicial complex $X$ ($N_0=225,N_1=629,N_2=405$) by treating each hexagon as a node, with edges for each pair of hexagons that share a side, and triangles for each set of three hexagons that share a corner.
This vector field is converted to a vector $\boldf\in\cC^1(X)$ by taking the flow on each edge to be the total flow perpendicular to the corresponding side between hexagons.
Then, dictionaries of joint and separated Hodgelets are constructed with $M=M_U=M_L=4$.
The joint dictionary has $2516$ atoms and forms a tight frame, while the separated dictionary has $5032$ atoms and also forms a tight frame.

We now consider the sparsity of the representation of $\boldf$ in each dictionary.
For each dictionary $\cD$ of atoms in $\cC^1(X)$, we construct the sparsest linear combination $\widehat{\boldf}$ of atoms in $\cD$ via orthogonal matching pursuit~\cite{Pati:1993,Tropp:2007} such that $\|\widehat{\boldf}-\boldf\|\leq\epsilon$, for $\epsilon\in(0,\|\boldf\|]$ sampled on a logarithmic scale.
As a baseline, we repeat this task using the Fourier basis vectors, \textit{i.e.,} the eigenvectors of the Hodge Laplacian, as well as with Fourier basis vectors using the linegraph Laplacian~\cite{Schaub:2018}, and wavelets constructed from the linegraph Laplacian.
The results of this are plotted in \cref{fig:results}~(b), where it is apparent that the separately designed dictionary outperforms the jointly designed one, with both performing better than the Fourier bases and the linegraph dictionary.

\noindent\textbf{Clustering buoy trajectories.}
One setting in which $1$-cochains find particular utility is in modeling flows and trajectories~\cite{Schaub:2018,Schaub:2020,Roddenberry:2021}.
We consider a dataset from the Global Drifter Program dataset, restricted to the region around Madagascar.\footnote{Data available from NOAA/AOML at \url{http://www.aoml.noaa.gov/envids/gld/}}
In this dataset, a set of buoys floating in the ocean have their location logged every 12 hours, which we use to construct trajectories along the edges of a triangulation ($N_0=133,N_1=320,N_2=186$) of the region.
Since these trajectories consist of a combination of oriented edges, they are naturally modeled as vectors in $\cC^1(X)$, following the approach taken by~\cite{Schaub:2020,Roddenberry:2021}.
We picture the sum of all $P=334$ such trajectories in \cref{fig:results}~(c), where the hole in the simplicial complex corresponds to the landmass of Madagascar.

We aim to find a good set of representative trajectories that captures both the local and global structure of the simplicial complex.
For a given dictionary, we first transform each trajectory by taking the inner product with each element of the dictionary.
Then, we perform sparse $k$-means~\cite{Witten:2010,sparcl} with $K=2$ clusters on $P_{tr}=0.75P$ trajectories in the dataset, which yields a set of centroids as well as feature selection weights.

To evaluate the quality of these centroids, we use the remaining $P_{ts}=0.25P$ trajectories as a test set, and compute the average normalized inner product between each test trajectory and the nearest centroid flow.
That is, for a set of centroids $\{\bc_k\}_{k=1}^K$ in $\cC^1(X)$, we compute for the test set $\{\boldf_j\}_{j=1}^{M_{ts}}$ the value
\begin{equation}
    L=\frac{1}{P_{ts}}\sum_{j=1}^{P_{ts}}\max_{1\leq k\leq K} \frac{\langle\boldf_j,\bc_k\rangle}{\|\boldf_j\|_2\|\bc_k\|_2},
\end{equation}
where the norms and inner products are taken with respect to the feature weighting kernel obtained from the sparse $k$-means procedure.
If a set of centroids yields a large value of $L$, indicating that the set of nearest flows aligns to each centroid well, we interpret it to be a realistic model for the trajectories.
We gather the results of this experiment when using the standard basis, the Fourier basis, the joint Hodgelet dictionary, and the separate Hodgelet dictionary in \cref{fig:results}~(d).
The wavelet dictionaries were constructed with $M=M^U=M^L=16$ filter banks.
We observe that the standard basis is not suitable for this task, since two trajectories that are close in space could have disjoint support, so the highly localized basis of edges does not perform well.
The Fourier basis and joint dictionary perform similarly, while the separate dictionary significantly outperforms the other representations.
This is due to the ability of the sparse clustering algorithm to both pick the proper scale for representation and discern between different qualitative features of the signal in this representation, as revealed by \cref{prop:div-curl-wavelets}.

\section{Conclusion}

Signals supported on the edges of simplicial complexes have been of great interest lately, with applications in computer graphics, mobility analysis, and modeling of physical flow phenomena.
We have considered the extension of spectral graph wavelets to such signals by replacing the graph Laplacian with the analogous Hodge Laplacian.
In doing so, we open up the possibility of considering the upper and lower components of the Hodge Laplacian separately, in order to yield a dictionary of wavelet atoms that respects the Hodge decomposition.
Based on these constructions, we state frame bounds for each type of wavelet dictionary, and then illustrate their utility for sparse representation on synthetic flow data.
Leveraging the ability of these wavelet dictionaries to sparsely represent flow signals, we demonstrate how they can be used to find high-quality representative cluster centroids on real-world buoy trajectory data via a sparse $k$-means procedure.

\ifdefined\isARXIVversion

\begin{appendices}

\crefalias{section}{appendix}

\section{Proof of Proposition 1}\label{app:div-curl}

Let $g_m^U,g_{m'}^L:\bbR^{>0}\to\bbR$ be given, such that $g_m^U(0)=g_m^L(0)=0$.
For any $j\in[N_1]$, we have
\begin{equation}
    \psi_{j,m}^U = g_m^U(\Delta_1^U)\be_j = g_m^U(\del_2\del_2^\top)\be_j.
\end{equation}
Let $\{(\bu_k,s_k,\bv_k)\}_{k=1}^{\rank\del_2}$ be the (low-rank) singular values and vectors of $\del_2$, so that
\begin{equation}
    \del_2 = \sum_{k=1}^{\rank\del_2}s_k\bu_k\bv_k^\top.
\end{equation}
Then, due to the fact that $g_m^U(0)=0$, applying the functional calculus yields
\begin{equation}
    g_m^U(\del_2\del_2^\top)\be_j = \sum_{k=1}^{\rank\del_2}g_m^U(s_k^2)\bu_k\langle\bu_k,\be_j\rangle.
\end{equation}
That is, $\psi_{j,m}^U$ can be written as a linear combination of the left (low-rank) singular vectors of $\del_2$, hence $\psi_{j,m}^U\in\Ima(\del_2)$.

A similar argument holds for the lower wavelets $\psi_{j',m'}^L$, completing the proof.\hfill$\blacksquare$

\section{Proof of Theorem 3}\label{app:tight-frame-sep}

Let $\boldf\in\cC^1(X)$ be given, and put
\begin{equation}
    S=\sum_{\psi\in\cD}|\langle\psi,\boldf\rangle|^2.
\end{equation}
This can be expanded to
\begin{equation}
    S = 
    \sum_{m=1}^{M_U}\sum_{j=1}^{N_1}|\langle g_m^U(\Delta_1^U)\be_j, \boldf\rangle|^2 
    + \sum_{m=1}^{M_L}\sum_{j=1}^{N_1}|\langle g_m^L(\Delta_1^L)\be_j, \boldf\rangle|^2.
\end{equation}
Since $\Delta_1^U,\Delta_1^L$ are Hermitian, this can be rewritten as
\begin{equation}
    \begin{aligned}
    S = 
    \sum_{m=1}^{M_U}\sum_{j=1}^{N_1}|\langle g_m^U(\Delta_1^U)\boldf, \be_j\rangle|^2
    + 
    \sum_{m=1}^{M_L}\sum_{j=1}^{N_1}|\langle g_m^L(\Delta_1^L)\boldf, \be_j\rangle|^2.
    \end{aligned}
\end{equation}

We now lower bound $S$.
Since the set $\{\be_j\}_{j=1}^{N_1}$ is an orthogonal basis for $\cC^1(X)$, Parseval's identity holds.
This yields
\begin{equation}\label{eq:first-lfb}
    S \geq 
    \sum_{m=1}^{M_U}\|g_m^U(\Delta_1^U)\boldf\|^2
    + 
    \sum_{m=1}^{M_L}\|g_m^L(\Delta_1^L)\boldf\|^2.
\end{equation}
Following the simplicial Fourier transform of \cite{Yang:2021}, let $(\mu_j,\bv_j)$ be the nonzero eigenpairs of $\Delta_1^U$ and $(\nu_j,\bu_j)$ be the nonnull eigenpairs of $\Delta_1^L$.
Then, there exists $\boldf^H\in\ker\Delta_1$ and real coefficients $\alpha_j,\beta_j$ such that
\begin{equation}
    \boldf = \boldf^H 
    + \sum_{j=1}^{\rank\Delta_1^U}\alpha_j\bv_j
    + \sum_{j=1}^{\rank\Delta_1^L}\beta_j\bu_j
\end{equation}
and
\begin{equation}
    \|\boldf\|^2 = \|\boldf^H\|^2
    + \sum_{j=1}^{\rank\Delta_1^U}\alpha_j^2
    + \sum_{j=1}^{\rank\Delta_1^L}\beta_j^2.
\end{equation}
With this in mind, coupled with the fact that $\boldf^H\in\ker(\Delta_1)=\ker(\Delta_1^U)\cap\ker(\Delta_1^L)$,
\eqref{eq:first-lfb} can be written as
\begin{equation}\label{eq:second-lfb}
    \begin{aligned}
    S &\geq 
    \left(\sum_{m=1}^{M_U}|g_m^U(0)|^2+\sum_{m=1}^{M_L}|g_1^L(0)|^2\right)\|\boldf^H\|^2 \\
    &\quad+
    \sum_{j=1}^{\rank\Delta_1^U}
    \Big(
    \sum_{m=1}^{M_U}|g_m^U(\mu_j)|^2
    +\sum_{m=1}^{M_L}|g_m^L(0)|^2
    \Big)
    \alpha_j^2 \\
    &\quad+
    \sum_{j=1}^{\rank\Delta_1^L}
    \Big(
    \sum_{m=1}^{M_U}|g_1^U(0)|^2
    +\sum_{m=1}^{M_L}|g_m^L(\nu_j)|^2
    \Big)
    \beta_j^2.
    \end{aligned}
\end{equation}
Considering our definition of the function $G$, \eqref{eq:second-lfb} can be more concisely written as
\begin{equation}
    \begin{aligned}
    S &\geq 
    G(0,0)\|\boldf^H\|^2
    +
    \sum_{j=1}^{\rank\Delta_1^U}
    G(\mu_j,0)
    \alpha_j^2 \\
    &\quad+
    \sum_{j=1}^{\rank\Delta_1^L}
    G(0,\nu_j)
    \beta_j^2.
    \end{aligned}
\end{equation}
One can then see that
\begin{equation}
    S\geq A\|\boldf\|^2,
\end{equation}
where $A$ is as defined in \eqref{eq:frame-bound-defn}.
A similar argument using the upper frame bounds yields
\begin{equation}
    S\leq B\|\boldf\|^2,
\end{equation}
as desired.

\end{appendices}

\fi

\newpage

\bibliographystyle{IEEEbib}
\bibliography{ref}

\end{document}